# A Low-Footprint Class Loading Mechanism for Embedded Java Virtual Machines

Christophe Rippert, Alexandre Courbot, Gilles Grimaud
`{Christophe.Rippert,Alexandre.Courbot,Gilles.Grimaud}@lifl.fr`
IRCICA/LIFL, Univ. Lille 1, UMR CNRS 8022, INRIA Futurs, POPS research group*

**Abstract**

This paper shows that it is possible to dramatically reduce the memory consumption of classes loaded in an embedded Java virtual machine without reducing its functionalities. We describe how to pack the constant pool by deleting entries which are only used during the class loading process. We present some benchmarks which demonstrate the efficiency of this mechanism. We finally suggest some additional optimizations which can be applied if some restrictions to the functionalities of the virtual machine can be tolerated.

## 1 Introduction

Embedding Java applications on resource-limited devices is a major ambition in a highly heterogeneous world where computing power is found in all kind of unusual devices. The portability of Java is an invaluable asset for the programmer who needs to deploy applications on these heterogeneous platforms. However, embedded Java virtual machines are typically very restricted because of the limitations of the underlying hardware. For instance, the JavaCard virtual machine [1] does not support dynamic class loading or garbage collection due to the typical computing power and memory space available on smart cards [2]. Memory is an especially scarce resource in most embedded systems due to technical constraints which prevent the miniaturization of large memory banks. Thus, reducing the size of the virtual machine and its runtime memory consumption are critical objectives if complex applications are to be executed on the system.

Reducing the memory space consumed by classes obviously means trying to obtain smaller code and smaller data. Previous work has shown that bytecode compression can be used to reduce the memory space used by the code [3], so it seems interesting to try and compress the data located in the constant pool of each class. A careful analysis of the constant pool shows that lots of its entries are only needed during the class loading process, and can be lost before execution. Thus, we have devised a new class loading mechanism which compacts the constant pool on-the-fly by suppressing entries as soon as they are deemed unnecessary. A valuable asset of our mechanism is that it does not imply disabling important functionalities of the virtual machine, such as dynamic type checking or garbage collection for instance.

We first present the JITS platform we have developed to build customized Java virtual machines for embedded systems. We then detail the class loading scheme we have chosen

*This work is partially supported by grants from the CPER Nord-Pas-de-Calais TACT LOMC C21, the French Ministry of Education and Research (ACI Sécurité Informatique SPOPS), and Gemplus Research Labs.

in JITS and present the optimizations we have implemented to reduce the memory space needed by loaded classes. Some evaluations of the memory consumption of various loaded classes are then presented, and we conclude by detailing the future optimizations we plan to implement in JITS.

## 2  JITS: Java In The Small

JITS is a Java-based operating system targeted at resource-limited devices, such as smartcards for instance. JITS goal is to provide a full-featured JVM and a complete API, which can then be customized to fit the needs of the applications and exploit at best the available resources. Therefore, by selecting only the packages needed by its applications, the developer can embed a fully tailored Java Runtime Environment without sacrificing functionalities of the virtual machine, as it is often the case in most embedded environments [2]. Indeed, many other attempts to embed Java on very small devices impose restrictions in their specification that requires using new languages or tools [1]. On the contrary, JITS doesn't impose restrictions from the specification level, but instead generates the ready-to-run ROM image containing both the customized Java-OS and applications to be run.

JITS is composed of a complete Java virtual machine, an API compliant with the Java 1 specification and a set of tools dedicated to help building the embedded JRE. These tools include the romizer, which is a standard Java program in charge of generating the binary image of the environment before it is loaded on the device. JITS can currently generate stand-alone virtual machines and ROM images running on ARM-based platforms. The compiled platform can also be used on the Intel ia32 architecture for debugging purposes.

The ROM generation process (romization) is the key part of the OS generation. It consists in loading all classes selected to be part of the embedded environment, and bringing them to an initialized state by following the standard *load-link-init* scheme [4]. The initialized classes are then "frozen" and dumped into a C file which will be compiled with the core of the virtual machine to build the runtime environment. The result of the compilation is a ROM image of the system. This ROM image is then burnt into the embedded device so the frozen system can continue its execution on the target device using the JITS VM. Thus, the romization is the process of bootstrapping and loading the OS on a system, and then suspending its execution and moving its image to another system so it can continue its execution where it stopped (Figure 1).

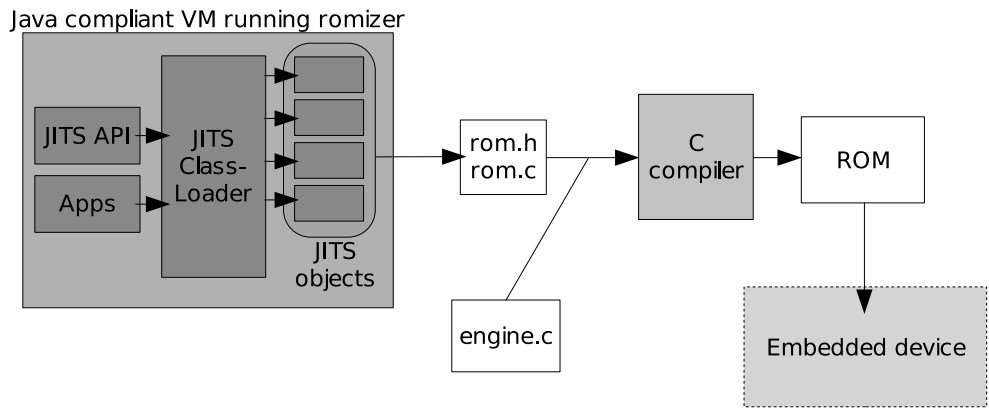

Figure 1: The romization process

The romizer can run on any JVM but it uses JITS classloader to create and initialize the JITS structures representing classes. Being able to run the romization process on any vir-tual machine differs from standard romization schemes which usually impose a dedicated building environment [5]. Similarly, the JITS API can be used as any other Java API by programs executed on a standard virtual machine.

This total compatibility with Java along with the usage of tools that work with `.class` files to produce the embedded system allow the programmer to develop and debug his embedded applications from the comfort of his favorite Java development tools. It is only once the applications are finalized that he uses JITS to produce the smallest possible ROM image of both the applications and the Java-OS to run them. He can also dynamically load his `.class` files into an already existing JITS VM, provided the VM has been built with dynamic loading support.

## 3  Class loading in JITS

Most classes loaded by JITS go through the four states presented below. Primitive types and arrays are exceptions to this scheme, since they are directly created by the virtual machine without having to load any class file. A class goes from state *unloaded* to state *loaded* after its `.class` file has been read and all references to internal fields and methods have been resolved (and optimized). It moves to state *linked* when all references to external entities (i.e. classes, fields and methods) have been resolved. Finally, it attains state *ready* when its static fields have been initialized. It is important to note that a class in state *loaded* can also be *ready*. A class become *ready* as soon as it can be used, whether the symbolic references have been resolved or not. It is still possible to link a class in state *loaded + ready* though, as shows figure 2.

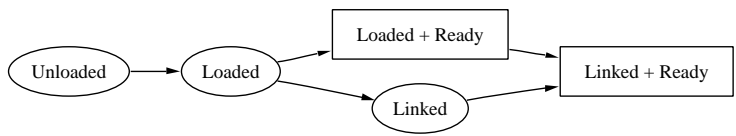

Figure 2: The lifecycle of a class in JITS (squares denote runnable states)

This scheme and the loading mechanism described below are the same whether the classes are romized or loaded dynamically after the device has been issued (i.e. the same `ClassLoader` is used in both cases).

### 3.1  State *unloaded*

A class is *unloaded* when its `Class` object is first created by a classloader (i.e. by using a `new Class()` instruction). This `Class` object is basically empty and the class still has to be loaded from its `.class` file. A class in state *unloaded* is basically a class which is referenced by another class but which has not yet been loaded by a classloader.

### 3.2  State *loaded*

A class is loaded when the `loadClass` method of a classloader is called. After having checked that the class has not already been loaded and having found its class file in the classpath, the classloader calls the `load` method of class `Class`. This method first reads the basic information of the class (i.e. its version number, name, superclass, etc) before loading its constant pool. When loading a class, JITS ignores attributes not useful for the execution of the program (e.g. line number table, source file, etc). This can save a significant memory space especially if the class file contains lots of debugging information.

The constant pool of classes is loaded from the `.class` file in two tables, named `atable` and `vtable`. The `atable` is an array of `Object` which is used to store `Utf8` constants, whereas the `vtable` is an array of `int` in which immediate values are encoded. The constant pool is then prelinked, which consists in resolving the accesses to some `Constant_info` structures. For instance, a `Constant_Class_info` is represented in the class file by a structure containing an index pointing to a `Constant_Utf8_info` entry in the constant pool. In JITS, a class constant is represented by a corresponding `Class` object stored in the `atable`. This `Class` object is later used to load the referenced class, which means that it would have been created anyway. Thus the `Constant_Utf8_info` entry can be suppressed if it is not referenced by any other entry. Similarly, the `Constant_String_info` structure is represented by an index in the `vtable` pointing to the string stored in the `atable` and the `Constant_NameAndType_info` structure is mapped as a couple of indexes to the `Utf8` stored in the `atable` and representing the name of the field or method and the descriptor of the type. A second pass of the prelinker transforms the `Constant_Fieldref_info`, `Constant_Methodref_info` and `Constant_InterfaceMethodref_info` entries into an `int` stored in the `vtable` and composed of the indexes of the corresponding `Class` object and `Field` or `Method` objects which are added to the `atable`. Thus, since the `Field` and `Method` objects would have been created when loading the fields and methods of the class, we are again able to save memory space by suppressing unused `Constant_Utf8_info` entries in the constant pool.

After loading the constant pool, the load method reads the interfaces implemented by the class, then its fields and its methods. The static fields of the class are stored in two tables, `aStaticZone` which contains reference fields, and `vStaticZone` for immediate values. Reading the methods consists in loading the bytecode, reading the exception table, loading stack maps if they are included in the class file, and finally building the class virtual method table. When loading the bytecode of a method, some instructions are replaced by an optimized version which will be interpreted faster at runtime and can also save some memory space. For instance, the `anewarray` instruction includes a constant pool index pointing to the type of the elements of the array. This instruction is replaced by `anewarray_quick`, which takes as a parameter an index pointing to an entry in the `atable` containing a `Class` object of the array component type. Thus, we can suppress the `Constant_Class` and `Constant_Utf8_info` entries representing the type of the elements of the array. Another interesting example of instruction replacement concerns the `ldc`, `ldc_w` and `ldc2_w` instructions. When loading the bytecode, these instructions are replaced by their *quick* counterparts which directly access the immediate value stored in the `vtable` without needing the `Constant_Integer_info`, `Constant_Float_info`, `Constant_Long_info` and `Constant_Double_info` structures. Thus, a `ldc` instruction is replaced by a `ldc_quick_a` instruction if the constant is a reference, a `ldc_quick_i` if the constant is an `int` and a `ldc_quick_f` if the constant is a `float`. It would be possible to use the same instruction for both `int` and `float` constants since they are both 32-bit immediate values, but that would compromise the type-checker which would not be able to differentiate `int` and `float`. By replacing `ldc` instructions by a type-specific opcode, we can preserve necessary type information without keeping complete constant pools entries and so preserve memory space.

## 3.3 State *linked*

Classes reach the *linked* state after being linked to each other. The linking process starts by recursively loading all the classes referenced by the constant pool of the class being linked. Then every method of the class is prelinked, which consists in type-checking its bytecode, compacting `invokevirtual` instructions and marking the constant pool entries used by the method code. During prelinking of a method, `invokevirtual` instructions are compacted if the index of the method in the constant pool and the number of arguments of the method are both lesser than 256. Compacting these instructions simply consists in replacing the index of the method in the constant pool, which is encoded on 16 bits in the instruction, by the number of arguments of the method and its index in the virtual method table of the class declaring it. Thus, at runtime the interpreter can call the method directly without accessing the constant pool, which speeds up the calling process. It also saves memory space since the constant pool entry representing the called method can be deleted. During method prelinking, constant pool entries which are used by the bytecode are marked so unused entries can be detected during the compaction of the constant pool.

Static fields referenced in the `vtable` are then converted to references pointing to the `vStaticZone` and `aStaticZone`. Static fields are treated differently than virtual fields since their value can be accessed directly in the `aStaticZone` and `vStaticZone`. The index pointing to the `Field` object representing the field is replaced by 16-bit immediate value containing the 13-bit offset of the field in the corresponding static zone and the 3-bit type of the field (which is necessary to know which static zone contains the field and how many bytes should be read). Thus, constant pool entries representing static fields can be suppressed.

The constant pool is then packed and resized, thereby loosing all unused entries. Finally, each method is linked, which basically means modifying the bytecode in order to replace indexes to the original constant pool entries by indexes to the corresponding compacted constant pool entries.

## 3.4 State *ready*

A class reaches the final state *ready* after initializing its static fields to their initial values. This is done in JITS by using the underlying virtual machine classloader to load the class and copy the values set by the static initializer to the JITS instance of the class. This rather heavy mechanism is necessary since the <clinit> method of a class cannot be called directly from a Java program.

## 4 Benchmarks

We monitored the memory footprint of JITS API when loaded using the scheme presented above. The API currently contains most classes from the base package `java.lang`, and some classes from `java.awt`, `java.io` and `java.net`, including a full TCP/IP stack.

We first monitored the number of constant pool entries discarded while loading the classes. Results are presented in Figure 3, with state *unloaded* refering to the number of entries in the `.class` files.

| Class state | *unloaded* | *loaded* | *linked* |
|---|---|---|---|
| Number of entries | 8,658 | 3,137 | 1,449 |
| % of initial number | 100% | 36.23% | 16.74% |

Figure 3: Number of constant pool entries for the whole JITS API

These results show that most of the reduction of the number of constant pool entries is done while loading the class, when resolving accesses to the constant pool and removing unnecessary indirections. We still manage to divide by two the number of entries while linking, by compacting `invokevirtual` instructions and packing static fields.

We then monitored the memory footprint of the constant pool in bytes. We tried and suppress as many `Constant_Utf8_info` as possible since they are the most space-consuming data in the constant pool. Unfortunately, some of them (e.g. field names, method descriptors, etc) are needed by the `java.lang.reflect` package so we need to keep them if we want to support introspection. Figure 4 presents the size of the constant pool with and without those strings to illustrate the cost of supporting introspection.

| Class state | unloaded | with introspection | | without introspection | |
|---|---|---|---|---|---|
| | | loaded | linked | loaded | linked |
| Size in bytes | 152,290 | 49,441 | 41,532 | 19,750 | 11,841 |
| % of initial size | 100% | 32.47% | 27.27% | 12.97% | 7.78% |

Figure 4: Size of the constant pool for the whole JITS API

The size of the constant pool can be lowered to less than 8% of its original size if introspection is not supported. This is due to the fact that direct references to `Constant_String_info` represent only a small part of all the `Constant_Utf8_info` constant pool entries, so most of them can be eliminated during loading. If these strings are not removed, we manage to pack the constant pool to nearly one fourth of its original size, while preserving a complete support for introspection.

Since most of our optimizations concern compacting of the constant pool (apart from disregarding unused attributes), we can use the results presented in Figure 4 to compute the size reduction for entire classes of JITS API. The total size of all `.class` files is 278,109 bytes[1], which includes 152,290 bytes for constant pools. Since we manage to reduce the size of constant pools to 41,532 bytes (with support for introspection), we finally obtain a memory footprint for the API which is only 60.17% of the total size of the `.class` files (Figure 5). It is interesting also to notice that dropping support for introspection saves us about 10% of the total `.class` file size.

| Class state | unloaded | with introspection | | without introspection | |
|---|---|---|---|---|---|
| | | loaded | linked | loaded | linked |
| Size in bytes | 278,109 | 175,250 | 167,351 | 145569 | 137660 |
| % of initial size | 100% | 63% | 60.17% | 52.34% | 49.5% |

Figure 5: Memory footprint of the whole JITS API

We finally compare our results with the *JEFF* class format [6]. *JEFF* is a proposal to replace the `.class` format by a more flexible, size-efficient, ready-to-run format which can include several classes into the same file with shared constant pools and indexed references to constant pool entries (instead of symbolic references as in `.class` files). JITS and *JEFF* are based on different approaches since with JITS, we chose to remains compliant with regular `.class` files which makes our virtual machine closer to Sun's specification.

The *JEFF* archive built from all the classes of JITS API is 186,766 bytes large, which is very similar to the 175,250 bytes we obtain with JITS when the classes reach state *loaded*.

---

[1]An important note concerns the total size of the API (nearly 280 KB). This might seem too much for an embedded virtual machine but it should be noted that JITS is a JavaOS which runs directly on the hardware. Thus, it does not need an underlying operating system, which saves some memory space. Also, unnecessary parts of this API for the target applications are not being included into the customized OS.

It has to be pointed out however that the size of *JEFF* structures once they are loaded should be slightly inferior to the file size (*slightly* because there is only little information to be discarded when loading a *JEFF* file, unlike the `.class` ones). It is foreseeable anyway that the two approaches could efficiently be combined. For instance, a *JEFF* file has a common constant pool for all the classes it includes, suppressing duplicate constant pool entries between classes, which JITS cannot do efficiently. A *JEFF* loader would allow JITS to take advantage of *JEFF*'s properties, reducing even more the memory footprint of loaded classes. Actually, a study of the constant pool of some classes reveals that many identical entries can be found between classes, like usual method prototypes (e.g. `()V` or `()I`). These entries are discarded if we don't require introspection, but a *JEFF* loader could bring our memory footprint closer to the "without introspection" results while still not degrading the system's features. The ideal regarding our philosophy would be to suppress duplicate constant pool entries while dynamically loading classes, but this would require doing things difficult to do in an embedded device, like relocating code if a `ldc` sees its index becoming bigger than 255 and needs to be turned into a `ldc_w`.

## 5 Future work

An optimization similar to the one applied to static fields can be done for private virtual fields. In JITS, objects are implemented as a C structure containing a pointer to the related class and the virtual fields of the object. When the `getfield` and `putfield` bytecodes are interpreted, the virtual machine accesses the required field by adding the offset stored in the bytecode to the base address of the object. Thus, it is possible to suppress all constant pools entries referencing private fields since all accesses to these fields are made in the class declaring them and so the `getfield` and `putfield` instructions can be modified to contain the proper offset. This optimization cannot be applied to protected, package-accessible or public fields since they could be accessed by a method of a class loaded dynamically after romization. In that case, the constant pool entries representing the fields would be necessary to link the new method.

However, if we define an additional state, called *package-closed*, we can apply this optimization to all non-public fields. The state *package-closed* is reached by classes in a package when no new class can be added to that package. Locking a package this way can be useful to prevent an application from modifying a fundamental package as `java.lang` for instance. If a class is *package-closed*, all constant pools entries corresponding to its non-public fields can be suppressed since all accesses can be linked before romization of the package.

Similarly, it is possible to define a state *closed* to be able to extend this optimization even to public fields. A class reaches the state *closed* if we can assure that no dynamically loaded class will need to be linked to this class. In practice, this state is most useful for embedded virtual machines romized with all the applications and that do not need to dynamically load new classes. These last two optimizations implies disabling some functionalities of the virtual machine (namely restricting or even forbidding dynamic class loading) so they will be made optional when implemented in JITS.

Preliminary results show that these optimizations would permit to reduce the constant pool below the 7.78% lower limit presented in Figure 4. In state *closed*, all `Constant_Utf8_info` representing name or type metadata would become useless, as well as all `Constant_Field_info` and `Constant_Method_info` entries in the constant pool. Thus, we can assume that 7.78% is the upper limit of the results we can expect when state *closed* is reached by a class, noting of course that closing a class prevent loading of any new class referencing it.

# 6   Conclusion

This paper shows that it is possible to greatly reduce the memory footprint of the class loading mechanism by applying on-the-fly packing of the constant pool of loaded classes. This allows us to save memory space on the embedded system without sacrificing functionality of the virtual machine, since for instance we can still type-check the bytecode of the class while suppressing type information from the constant pool, or use stack maps included in the class file (or dynamically generated by the type checker) to guide the garbage collector. Coupled with the flexibility of the JITS romization process, this permits to generate a Java virtual machine fully tailored for the target device, thus exploiting at best the limited resources available.